\documentclass{article}

\usepackage{arxiv}

\usepackage[utf8]{inputenc} % allow utf-8 input
\usepackage[T1]{fontenc}    % use 8-bit T1 fonts
\usepackage{hyperref}       % hyperlinks
\usepackage{url}            % simple URL typesetting
\usepackage{booktabs}       % professional-quality tables
\usepackage{amsfonts}       % blackboard math symbols
\usepackage{nicefrac}       % compact symbols for 1/2, etc.
\usepackage{microtype}      % microtypography
\usepackage{lipsum}
\usepackage{graphicx}
\usepackage{amsmath}
\usepackage{tabularx}

\title{Ultrafast laser pulse characterization by THG d-scan using optically enhanced graphene coatings}

\author{
  Tiago Gomes\\
  IFIMUP and Dept. of Physics and Astronomy\\
  Faculty of Sciences, University of Porto\\
  Rua do Campo Alegre s/n, 4169-007 Porto, Portugal \\
  \texttt{tsgomes@fc.up.pt} \\
   \And
  Miguel Canhota \\
  IFIMUP and Dept. of Physics and Astronomy\\
  Faculty of Sciences, University of Porto\\
  Rua do Campo Alegre s/n, 4169-007 Porto, Portugal \\
   \And
  Bohdan Kulyk\\
  i3N-Aveiro, Dept. of Physics\\
  University of Aveiro\\
  Aveiro, Portugal\\
 \AND
  Alexandre Carvalho\\
  i3N-Aveiro, Dept. of Physics\\
  University of Aveiro\\
  Aveiro, Portugal\\
 \AND
  Bruno Jarrais\\
  REQUIMTE/LAQV, Dept. of Chemistry and Biochemistry\\
  Faculty of Sciences, University of Porto\\
  Rua do Campo Alegre s/n, 4169-007 Porto, Portugal \\
 \AND
  António José Fernandes\\
  i3N-Aveiro, Dept. of Physics\\
  University of Aveiro\\
  Aveiro, Portugal\\
 \AND
  Cristina Freire\\
  REQUIMTE/LAQV, Dept. of Chemistry and Biochemistry\\
  Faculty of Sciences, University of Porto\\
  Rua do Campo Alegre s/n, 4169-007 Porto, Portugal \\
 \AND
  Florinda Costa\\
  i3N-Aveiro, Dept. of Physics\\
  University of Aveiro\\
  Aveiro, Portugal\\
 \AND
   Helder Crespo \\
   Blackett Laboratory \\
   Imperial College \\
   London SW7 2AZ, UK \\
}

\begin{document}
\maketitle

*Corresponding author: tsgomes@fc.up.pt

\begin{abstract}
We have successfully functionalized 5 layers of TCVD-grown graphene by photoassisted transfer hydrogenation reaction with formic acid, in order to increase its nonlinear optical response and laser-induced damage resilience. Using the functionalized graphene samples, we were able to fully characterize sub-10 fs ultrashort laser pulses using THG dispersion-scan, with results that are in excellent agreement with the ones obtained with  pristine graphene samples.
\end{abstract}

% keywords can be removed

\section{Introduction}
Graphene consists of a single atomic layer of carbon atoms laid out in an hexagonal lattice and is a promising material for many ultrafast photonics applications \cite{app1,app2}. Its high nonlinear third-order optical susceptibility \cite{nonlinear_response} allows for intense and broadband third-harmonic generation using femtosecond laser pulses at relatively low intensities, with the promising possibility of third-harmonic enhancement by using multi-layer graphene \cite{MIKHAILOV2012924}. 

However, these low intensities are still sufficient to interact with graphene's lattice \cite{Damage_threshold_graphene2}, creating localized heating that leads to  melting, vaporization, and/or sublimation \cite{Damage_threshold_graphene}. The possibility of enhancing graphene's nonlinear optical response, as well as its resilience to laser-induced damage, would certainly provide a solution to overcome this problem. Several techniques have already been studied in order to improve graphene's nonlinear optical response, like embedding it on dieletric ressonant waveguide gratings \cite{ZHAO201930}, or using graphene metasurfaces \cite{Jin_2017}.

In this Letter, we present a solution, based on photoassisted hydrogenation, to improve graphene's laser-induced damage resilience, as well as increasing its nonlinear optical susceptibility. By measuring the third-harmonic signal as a function of the dispersion applied to the pulse, we get a d-scan trace, that allows to retrieve the spectral phase of the used ultrashort pulses via retrieval algorithms \cite{Miranda:12,Miranda:13}. 

\section{Graphene Synthesis}
Graphene was synthesized by Thermal Chemical Vapor Deposition (TCVD), according to a process based on the one previously reported by Kulyk et al. \cite{KULYK2020403}. Briefly, 25 $\mu m$-thick copper foils ($>$99.99\%, MTI) were cut into ~2x4 $cm^2$ sheets and washed in acetone and isopropanol (15 min ultrasonication in each of the solvents). This substrate was placed inside the reactor, pre-heated to a temperature of 940 °C, and subjected to a 10 min annealing under 190 sccm of $H_2$ and 190 sccm of Ar, at 276 mbar. Next, the chamber was pumped down to ~0.13 mbar and ~200 sccm of air was introduced, for 5 min. Afterwards, the system was pumped down once again to ~0.15 mbar and 50 sccm of $H_2$ were introduced, for 20 min. The gas was then pumped out and the chamber was brought up to atmospheric pressure using Ar. Next, the reaction chamber was pumped down to 0.3 mbar. The substrate was then subjected to a $25\,\text{ºC/min}$ temperature ramp from $940\,\text{º C}$ to $1080\,\text{º C}$, under $190\,\text{sccm}$ of $H_2$ and $190\,\text{sccm}$ of Ar, followed by a $10\,\text{min}$ annealing at $1080\,\text{C}$ in the same atmosphere. The pressure during these two steps was allowed to rise to $276\,\text{mbar}$ and was then maintained at this value.

In the following stage, $CH_4$ was introduced into the chamber, alongside 38 sccm of $H_2$ and 200 sccm of Ar. The flowrate of $CH_4$ was set at 0.15 sccm for the first 6 min, followed by 1 min without any $CH_4$ supply (but maintaining the same flowrates of $H_2$ and Ar), and finally 40 min of 0.10 sccm of $CH_4$. The sample was then rapidly pulled out towards the cold end of the reactor and all the gases were pumped out, followed by the pressurization of the chamber to atmospheric pressure with Ar.

\begin{figure}
    \centering
    \includegraphics[width=0.7\linewidth]{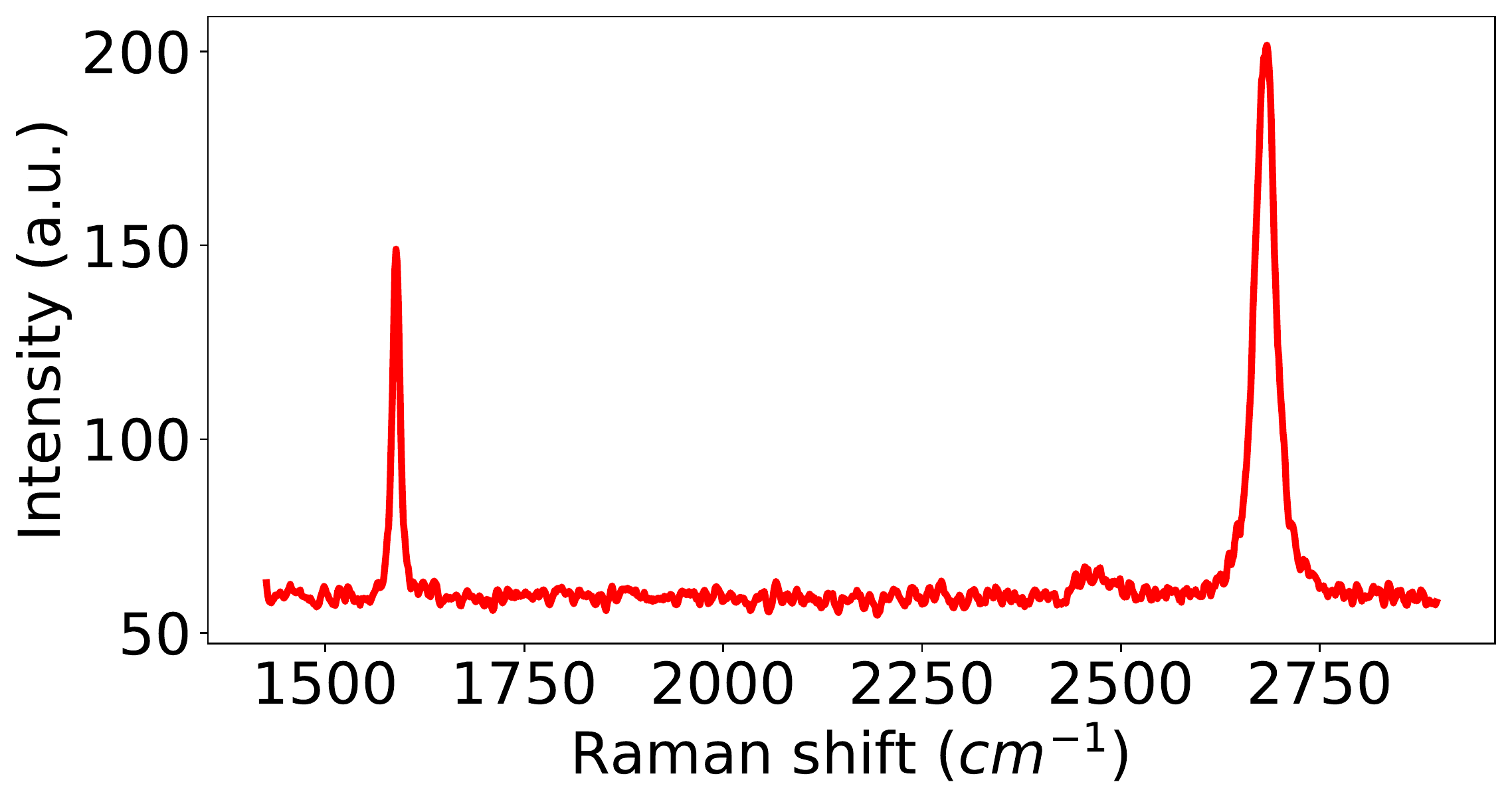}
    \caption{Raman spectra of the grown single-layer graphene sample.}
    \label{fig:Raman}
\end{figure}

A Raman spectrum for the single-layer graphene can be seen in Figure \ref{fig:Raman}. Raman spectroscopy shows the characteristic 2D and G peaks of graphene, with a large intensity ratio between them $\left(I_{2D}/I_G\approx1.5\right)$. Since the peaks are narrow and symmetrical, and the 2D peak (located at $2683.58\,cm^{-1}$) is more intense than the G peak (located at $1588.84\,cm^{-1}$), we can conclude that the grown sample is single-layer \cite{Ferrari2013RamanSA,Raman2,FERRARI200747}.

The as-grown single-layer graphene sheets were then successively transferred onto a fused silica substrate (1 mm-thick, University Wafer, grade JGS2), forming a stack of 5 layers of graphene. Each transfer was made by the widely used electrochemical bubbling technique, with PMMA, (average molecular weight 550 000, Alfa Aesar, 4.5 wt.\% in anisole) as a supporting polymer \cite{CARVALHO201699}.

\section{Graphene Functionalization}
For the functionalization process, the sample was then placed in a flat bottom flask. A 50 mL 1/1 (v/v) mixture of $HCOOH$ and $H_2O$ was added. After degassing the solution with a $N_2$ stream for 15 minutes, the setup was exposed to illumination in $Q-SUN$ $Xe-1$ xenon arc chamber fitted with a daylight filter during 94 hours. Afterwards, the substrate was removed from the solution and was thoroughly washed by rinsing with 100 mL ultrapure $H_2O$, and dried in an oven at $110\,\text{º C}$, under vacuum.

%__________________________Functionalization scheme______________________
\begin{figure}[htbp]
\centering
\fbox{\includegraphics[width=0.5\linewidth]{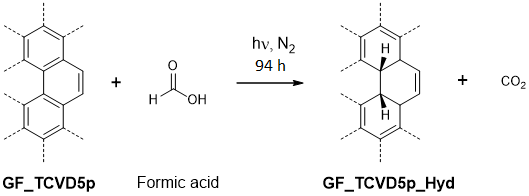}}
\caption{Preparation of hydrogenated graphene through photoassisted transfer hydrogenation reaction with formic acid.}
\label{fig:graphene_hydrogenation}
\end{figure}

For the functionalization characterization, X-ray photoelectron spectroscopy (XPS) was performed in a VG Scientific ESCALAB 200A spectrometer, using non-monochromatized Al $K\alpha$ radiation (1486.6 eV). To correct possible deviations caused by electric charge of the samples, the C1s band at 284.6 eV was taken as internal standard \cite{handbookXPS,paper2_XPS}. The XPS spectra were deconvolved with the CasaXPS software, using non-linear least squares fitting routine after a Shirley-type background subtraction. The surface atomic percentages were calculated from the corresponding peak areas and using the sensitivity factors provided by the manufacturer.

\begin{figure}[htbp]
\centering
\fbox{\includegraphics[width=0.8\linewidth]{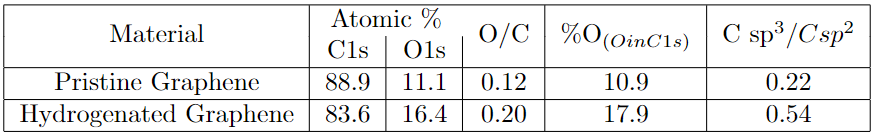}}
\caption{XPS surface atomic percentages for the pristine and hydrogenated TCVD graphene materials.}
\label{fig:table1}
\end{figure}

XPS analysis was carried out to assess the type and relative amount of functional groups in the pristine and functionalized (from now on denominated hydrogenated) TCVD graphene materials. The relative amounts of the different elements were calculated from the corresponding peak areas and are shown in Fig. \ref{fig:table1} for the pristine and hydrogenated TCVD graphene materials, respectively.

\begin{figure}[htbp]
\centering
\fbox{\includegraphics[width=0.8\linewidth]{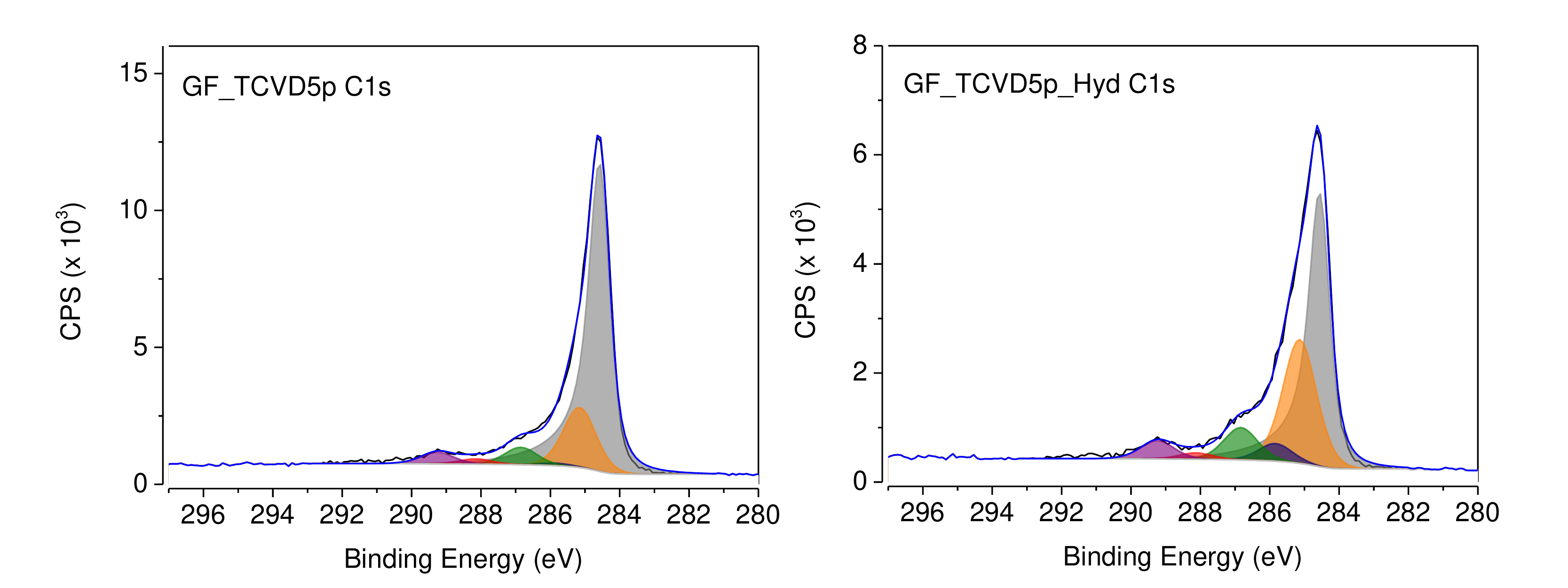}}
\caption{High resolution XPS spectra in C1s region of pristine and hydrogenated graphene.}
\label{fig:XPS}
\end{figure}

As can be seen in Fig. \ref{fig:table1}, there is an increase in the O relative atomic percentage upon funcionalization, and although it should be noted that the O1s high resolution spectra of these materials are non-reliable, due to the XPS sampling depth together with the fact that the samples are supported on fused silica substrates, the obtained O1s percentages are in agreement with the ones obtained from the C1s high resolution spectra: \%O (O in C1s). These percentages, as well as the C $sp^3$ / C $sp^2$ ratio were obtained from the deconvolution of the C1s high resolution spectra, shown in Fig. \ref{fig:XPS}.

\begin{figure}[htbp]
\centering
\fbox{\includegraphics[width=0.8\linewidth]{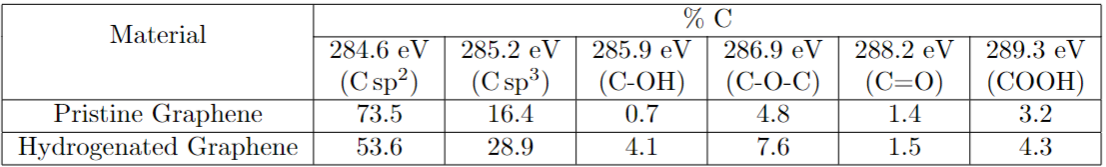}}
\caption{Relative atomic percentages of carbon-containing groups from the deconvolution of the C1s high resolution XPS
spectra of the pristine and hydrogenated TCVD graphene materials.}
\label{fig:table2}
\end{figure}

The spectra were fitted using an asymmetric line-shape for the $sp^2$ component and a symmetric one for the other components. All groups were assigned as follows: $sp^2$ regions (284.6 eV), $sp^3$ and other topological defects (+0.6 eV), hydroxyl (C-OH, +1.3 eV), epoxy (C-O, +2.3 eV), carbonyl (C=O, +3.6 eV), and carboxyl (COOH, +4.7 eV) \cite{XPS3}, and the results are presented in Fig. \ref{fig:table2}, for the pristine and hydrogenated TCVD graphene materials, respectively.

It is quite clear from Fig. \ref{fig:XPS} and \ref{fig:table2}, that there is a substancial increase in the C $sp^3$ component in hydrogenated graphene upon the photoassisted transfer hydrogenation reaction, at the expense of C $sp^2$ relative percentage, which indicates that the reaction was effective in transforming $sp^2$ carbons from the basal plane of graphene into $sp^3$ hydrogenated carbons.

\begin{figure}[htbp]
\centering
\fbox{\includegraphics[width=0.6\linewidth]{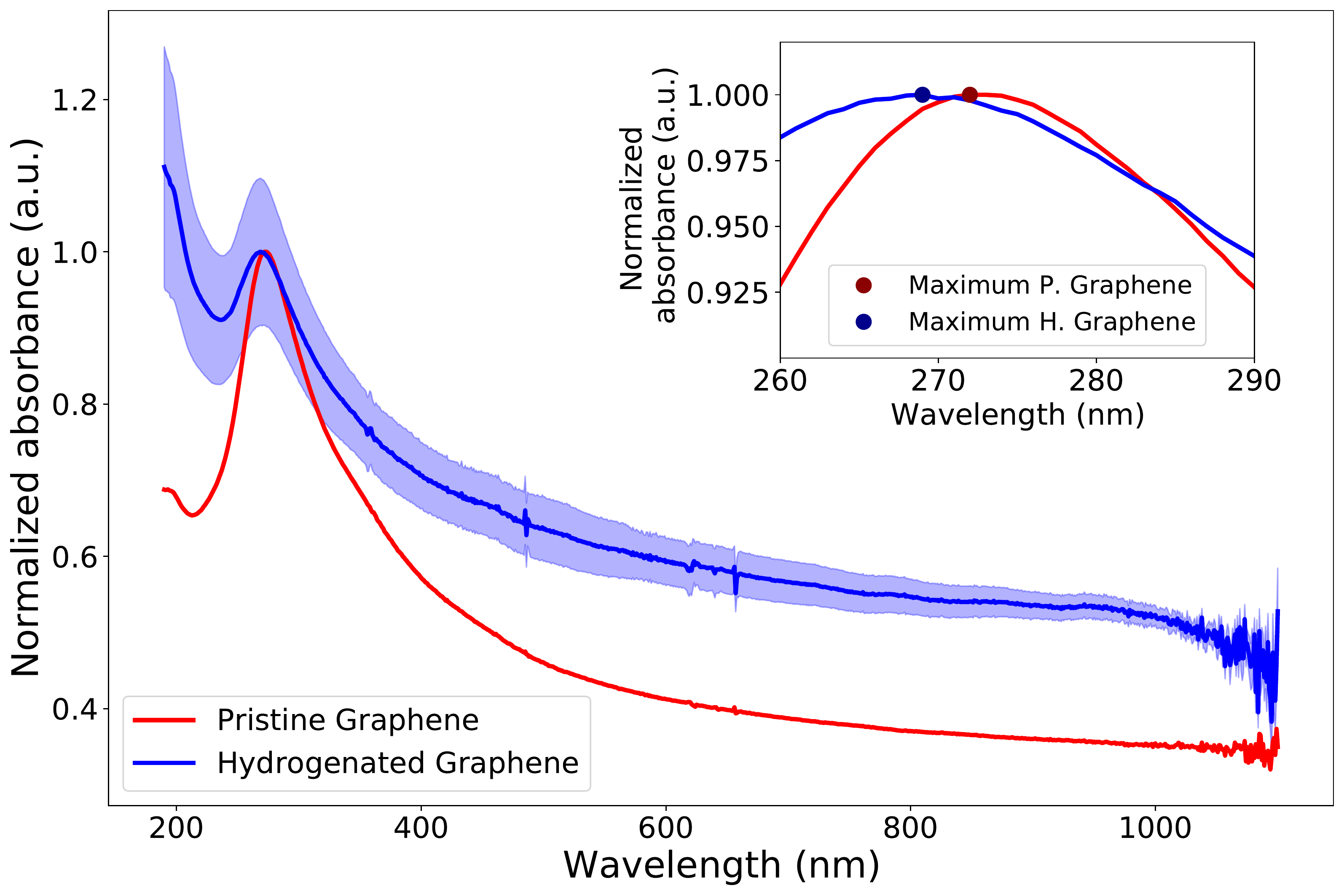}}
\caption{Absorption curves of 5 layers pristine and hydrogenated graphene.}
\label{fig:Abs_spectra}
\end{figure}

Fig. \ref{fig:Abs_spectra} depicts the normalized absorption spectra of the pristine and the hydrogenated graphene (averaged from 10 different zones). The absorption spectra were monitored in an Agilent 8453 UV-Visible spectrophotometer, in the range 190 – 1100 nm, with a 1 nm resolution. It is possible to identify the main absorption peak, $\lambda_{max}$, at $\approx 270\,\text{nm}$, attributed to $\pi\longrightarrow \pi^*$ transitions from the aromatic rings. However, it is also possible to observe (inset) a blue shift in $\lambda_{max}$, from $273$ to $268\,\text{nm}$, when going from the pristine to the hydrogenated material. This is due to the introduction of surface defects, namely the transformation of $sp^2$ hybridized carbon into $sp^3$, as the degree of aromatic conjugation can be determined by the $\lambda_{max}$ of the absorption spectra: the lower  $\lambda_{max}$ observed for the hydrogenated graphene means that more energy is needed for the electronic transition, as there are less $\pi\longrightarrow \pi^*$ transitions (conjugation) \cite{abs1,abs2}. This result corroborates the XPS observations regarding the transformation of $sp^2$ into $sp^3$ carbon upon the photoassisted hydrogenation.

\section{Third-Harmonic Generation Dispersion-scan}

\begin{figure}[htbp]
\centering
\fbox{\includegraphics[width=0.7\linewidth]{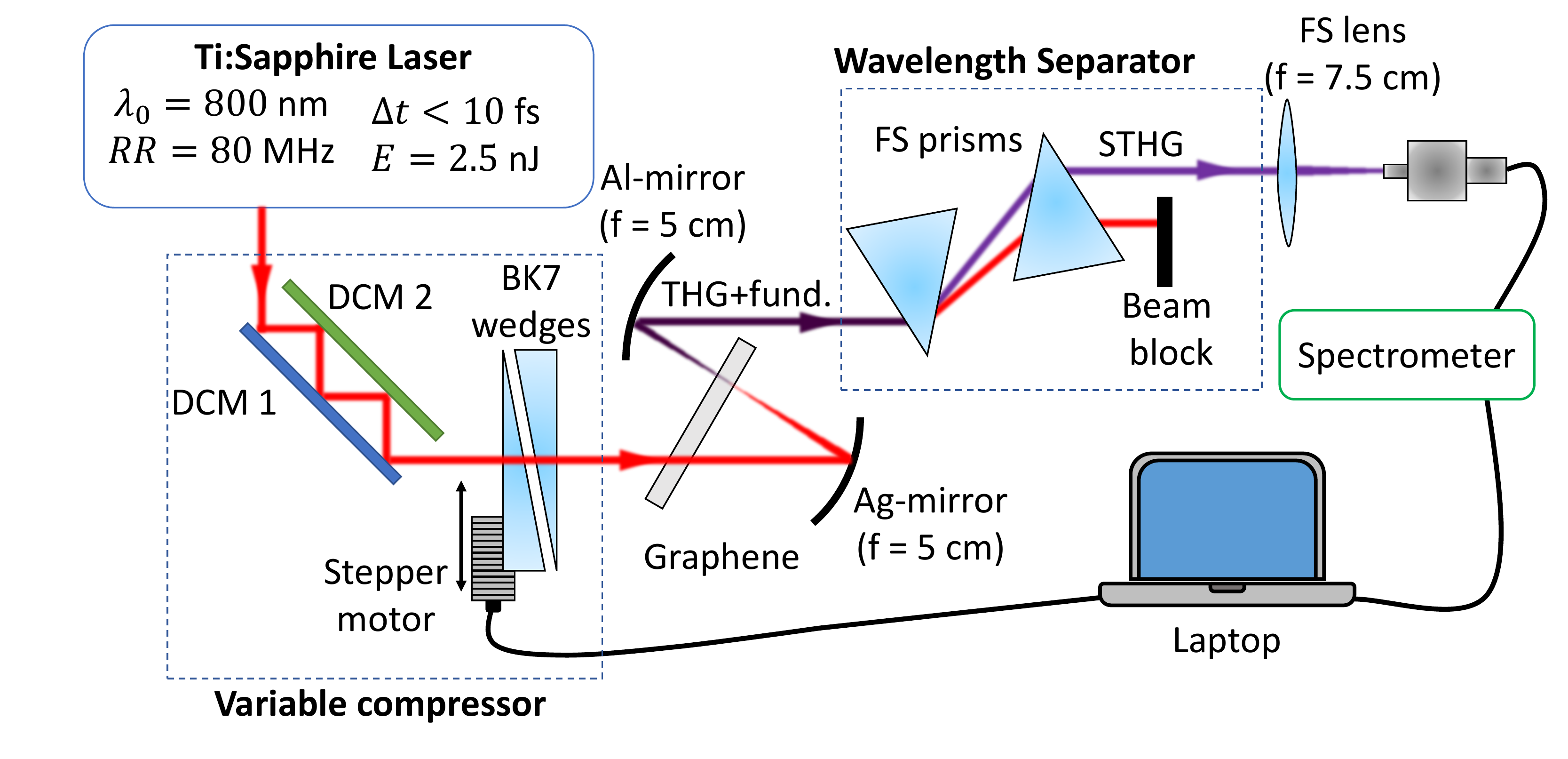}}
\caption{THG d-scan experimental setup. DCM1,2: double- chirped mirrors (see text for more details).}
\label{fig:THG_d_setup}
\end{figure}

To study the functionalization effect on the samples' third-order nonlinear optical susceptibility, a THG d-scan setup was developed, which is shown in Figure \ref{fig:THG_d_setup}. In our experiments, we used an ultra-broadband Ti:Sapphire laser oscillator (Femtolasers Rainbow CEP) delivering pulses with a central wavelength of $800\,\text{nm}$, sub-10-fs duration, $2.5\,\text{nJ}$ of energy and a repetition rate of $80\,\text{MHz}$. Like in most d-scan implementations, the setup comprises three sections: a variable compressor (usually already in place and responsible for adjusting the total dispersion applied to the laser pulse), a nonlinear signal generator (where the THG is performed) and a spectral measurement. In the variable compressor, the beam undergoes 8 bounces off ultra-broadband double-chirped mirrors (Laser Quantum Ltd) that introduce negative dispersion before crossing a pair of BK7-glass wedges for variable dispersion compensation. One of the wedges is connected to a motorized stage for scanning and fine-tuning the applied dispersion. The pulse energy is adjusted with a variable neutral-density filter prior to the nonlinear section, where the beam crosses the graphene sample and is focused back on the same sample with a concave silver-coated spherical mirror ($\text{f}=5\,\text{cm}$) at a incidence angle of $19^{\circ}$ (a smaller angle would have been preferable to minimize astigmatism, but this was the best compromise in our setup given our beam size and the useful aperture of the components). The peak intensity at the focus is of the order of $900 \, \text{GW.cm}^{-2}$. The resulting THG signal is then collimated with an aluminum-coated concave spherical mirror with the same focal length. Spherical mirrors were used instead of lenses to reduce chromatic aberration. Note that high-quality, low scatter off-axis parabolic mirrors could have been used instead. We opted for spherical mirrors at a relatively low incidence angle due to their better quality focused spot and higher focused intensity compared to standard off-axis parabolic mirrors. 

The colinear fundamental and THG beams are sent through a wavelength separator composed of a pair of prisms, to spatially separate the two beams prior to the spectral measurement \cite{silvaMidIRUltrabroadbandThird2013a}. A double-pass scheme was used to remove spatial chirp (for clarity, only one pass is shown in Figure \ref{fig:THG_d_setup}). The spectrum of the THG signal was then recorded as a function of dispersion with a fiber-coupled spectrometer (Ocean Optics HR4000) to obtain the measured THG d-scan trace. We scanned the dispersion in 150 steps over a 4-mm wedge insertion range for both samples, using an integration time of 500 ms.

\begin{figure}[htbp]
\centering
\fbox{\includegraphics[width=0.6\linewidth]{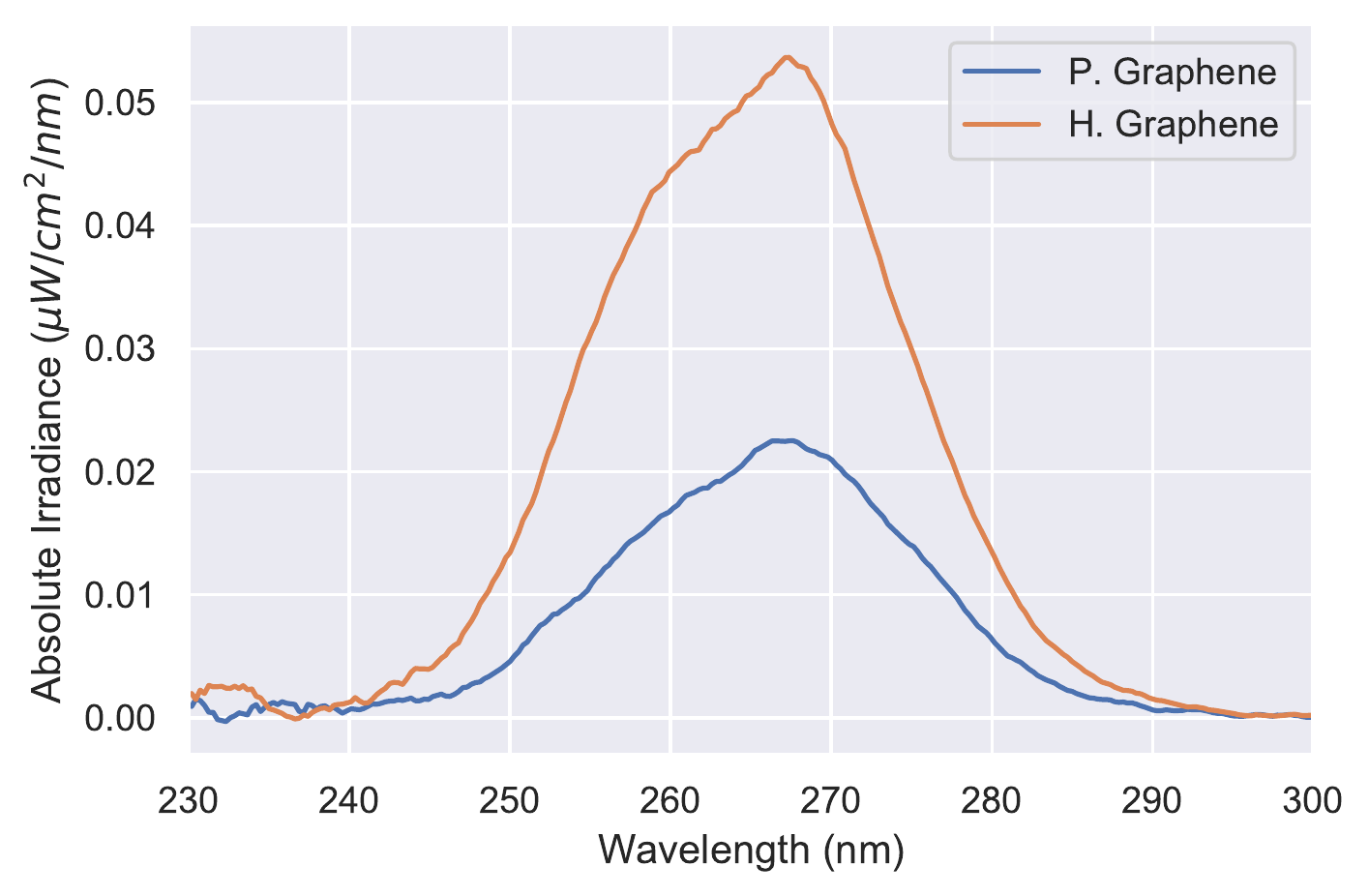}}
\caption{Third-harmonic signal comparison for hydrogenated (orange) and pristine (blue) TCVD-grown 5 layers graphene.}
\label{fig:THG_comp}
\end{figure}

Figure \ref{fig:THG_comp} shows the third-harmonic signal measured for TCVD-grown 5 layers graphene, both hydrogenated and pristine. It is clear that the functionalization process led to an increase in the THG efficiency, since the THG signal suffered an increase of almost 2.5x compared to the signal obtained with the pristine graphene sample. For lower wavelengths, the THG signal seems to have increased its SNR, corresponding to an increase in the spectrum's FWHM. However, in general terms, the structural shape of the third-harmonic signal has not changed.

%_________________________THG Generation Efficiency___________

\begin{figure}[htbp]
\centering
\fbox{\includegraphics[width=0.7\linewidth]{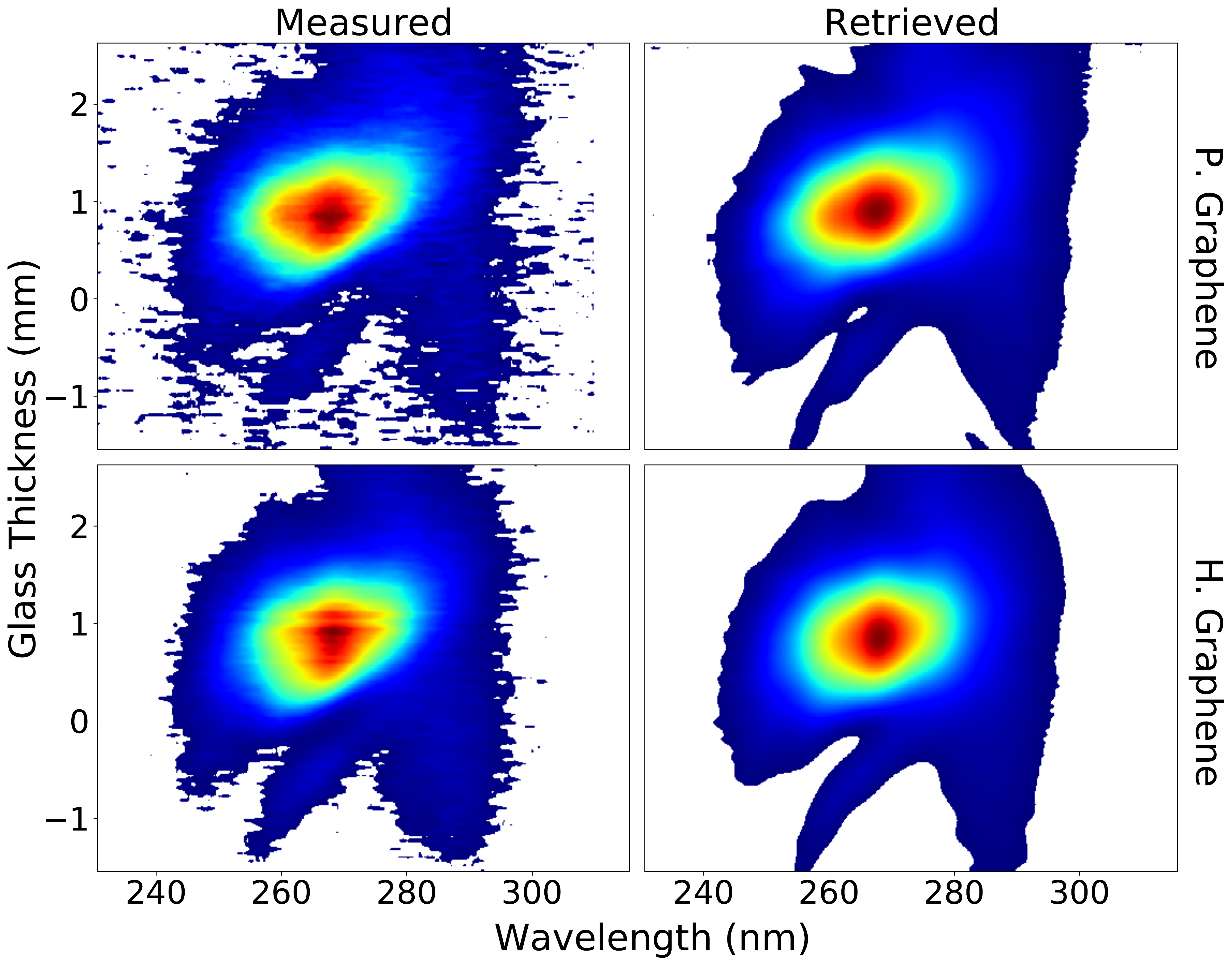}}
\caption{Measured (left) and retrieved (right) THG d-scan using the pristine (top) and hydrogenated (bottom) graphene samples.}
\label{fig:THG_dscan}
\end{figure}

By performing a d-scan measurement with both the hydrogenated and the pristine TCVD-grown 5 layers graphene sample (see Figure \ref{fig:THG_dscan}), we see that the SNR increased dramatically when generating third-harmonic with hydrogenated graphene. In both cases, a small tilt can be seen on the main trace, resulting from residual third-order dispersion. 

Following the d-scan measurements, an amplitude and phase retrieval was performed for both measurements. The retrieved d-scan traces shown correspond to one out of 10 retrievals, performed for different initial guesses (Gaussian spectra with random second- and third-order spectral phases). The retrieved spectral phase and intensity used to reconstruct the pulse's temporal profile are the mean retrieved spectral phase and intensity of all retrievals. 

Both retrievals show that the main features of the measured d-scan traces, namely the traces' tilt and the small features that appear around the traces center of mass, are very well reproduced. The retrieved traces are similar to one another, showing that the functionalization process only increased the absolute THG efficiency.

\begin{figure}[htbp]
\centering
\fbox{\includegraphics[width=0.6\linewidth]{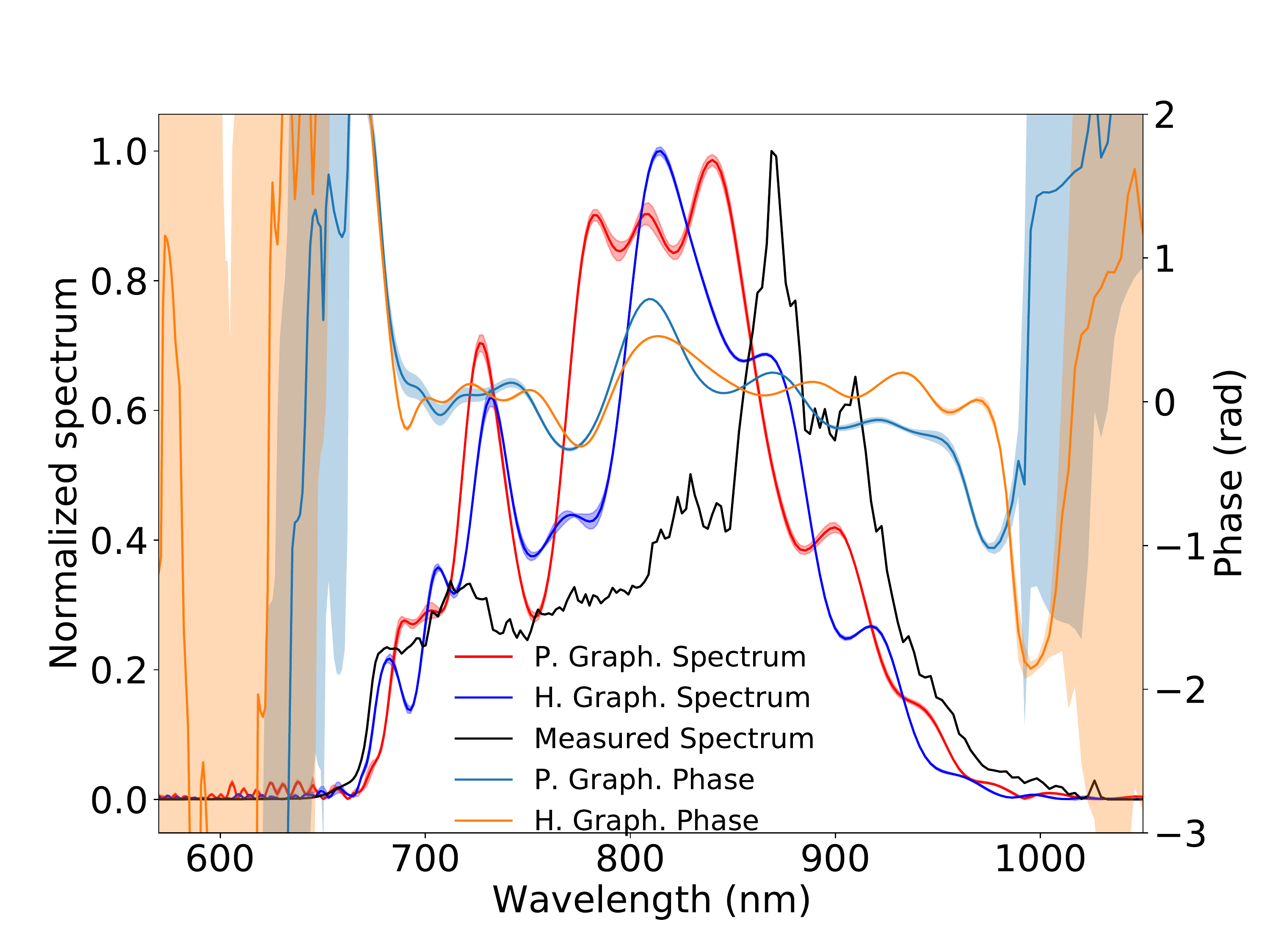}}
\caption{Mean retrieved spectral phase and amplitude for THG d-scan using pristine (light blue and red, respectively) and hydrogenated (orange and dark blue, respectively), along with the calculated standard deviation (shaded area). The black curve is an independent measurement of the spectral power of the laser.}
\label{fig:THG_specretr}
\end{figure}

By retrieving the spectral phase and amplitude for both processes, we get the results from Figure \ref{fig:THG_specretr}. 
Both retrievals led to similar spectral phases, with small differences in the zones with lower spectral power. The phase uncertainty in both cases is very low on the region with high spectral power. Both spectral phases diverge rapidly for low $\left(<680\,\text{nm}\right)$ and high $\left(>900\,\text{nm}\right)$ wavelengths, and are relatively flat in the $680-900\,\text{nm}$ region, meaning that the pulses are well compressed. A small third-order spectral phase component can be seen in both retrieved phases, which is in good agreement with the traces' small tilt.

The retrieved spectral power in both cases is similar to one another and to the measured spectrum.

% WHat does "P. Graph" and "H. Graph" means? Pristine and Hydrogenated. The first reference is on the absorption spectra

\begin{figure}[htbp]
\centering
\fbox{\includegraphics[width=0.6\linewidth]{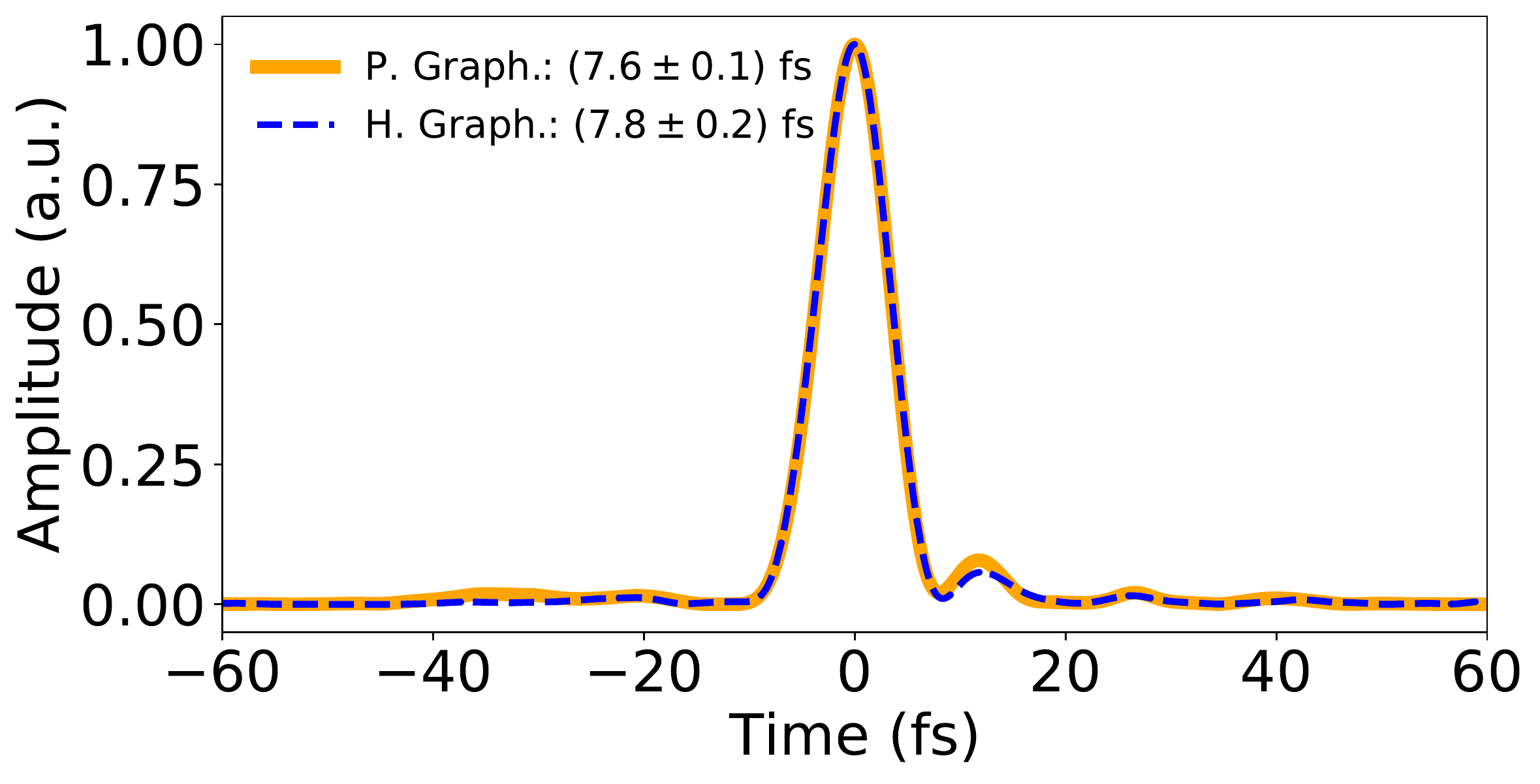}}
\caption{Pulse temporal profile retrieved by THG d-scan for the pristine (orange) and hydrogenated (blue) graphene samples, using both the retrieved spectral amplitude and phase.}
\label{fig:THG_temp_prof}
\end{figure}

The temporal profile for both processes are shown in Figure \ref{fig:THG_temp_prof}. Both temporal profiles have similar behaviour: a main pulse followed by a weaker post-pulse, resulting once again from residual third-order dispersion. The retrieved temporal profiles are in good agreement with each other and with previous measurements obtained with the same laser source \cite{Miranda:13}.

\begin{figure}[htbp]
\centering
\fbox{\includegraphics[width=0.6\linewidth]{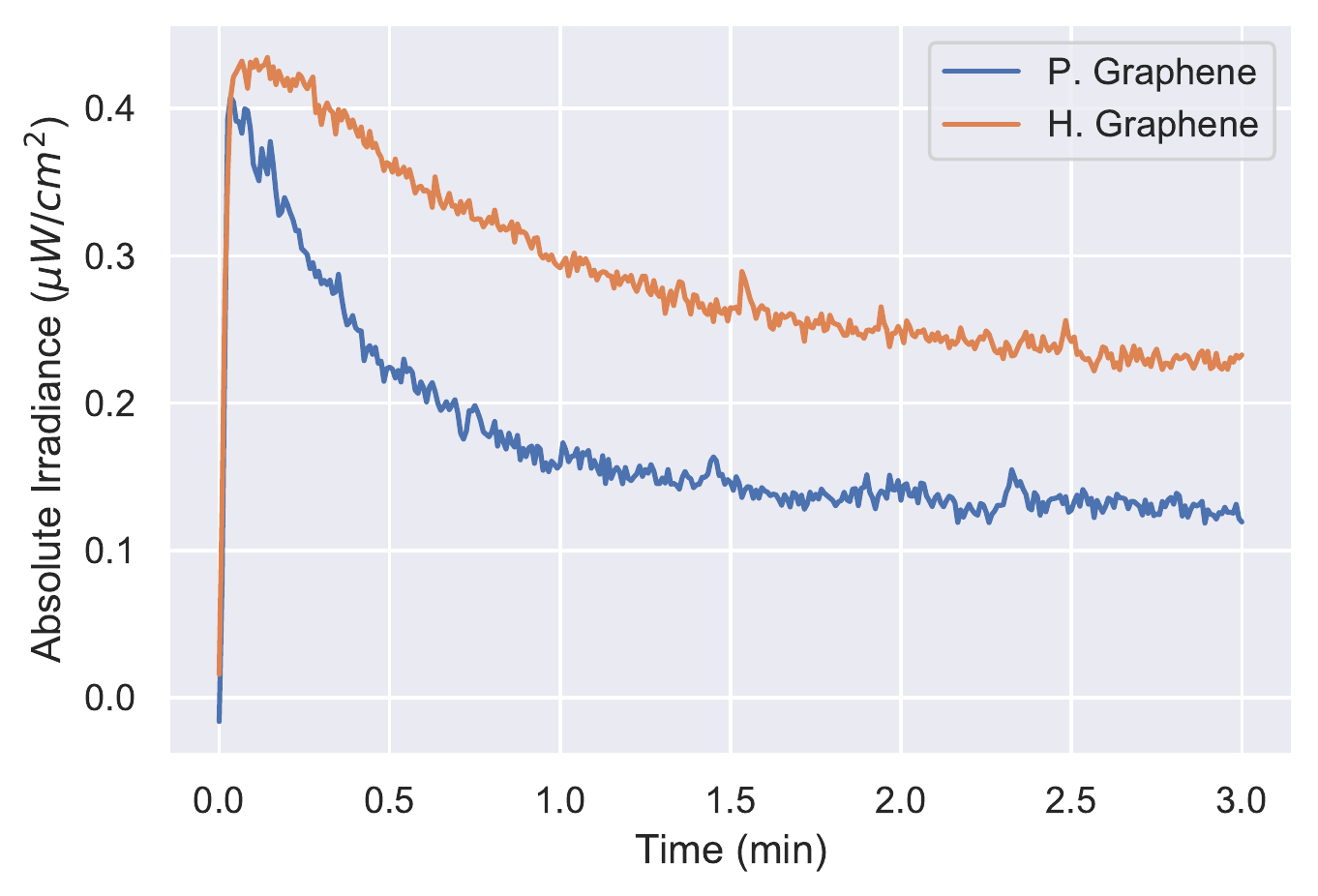}}
\caption{Exponential decay of the third-harmonic signal for the pristine (blue) and the hydrogenated (orange) graphene sample.}
\label{fig:durability}
\end{figure}

To study the effect of the functionalization process on the graphene's damage threshold, measurements of the graphene's integrated third-harmonic signal over time were performed. The results for both the hydrogenated and pristine graphene samples are shown in Figure \ref{fig:durability}.

Several measurements were made in different points of the samples, to determine the average decay rate of the third-harmonic signal. It is important to note that the measurements were made with the same third-harmonic power, which was achieved by placing the hydrogenated graphene sample slightly out of focus. The results show that the decay rate for the hydrogenated sample was reduced by 30\%, indicating that the functionalization improved the sample's resilience to laser-induced damage. Although the hydrogenated graphene sample was not subjected to the same intensity as the pristine graphene sample, which can be an explanation of this damage reduction, it is clear that, with the same third-harmonic power - that is already sufficient for the complete ultrashort pulse characterization - the decay of the third-harmonic power is slower.

A possible explanation for the damage caused to the graphene's sample is the strong DUV absorption. As already stated in Figure \ref{fig:Abs_spectra}, pristine graphene has an absorption peak at around 270 nm, which is very close to the third-harmonic center wavelength, therefore being a possible explanation of the graphene's degradation over time.

\section{Conclusion}

Single-layer graphene sheets were synthetized by TCVD on copper substrates. The as-grown sample was then successively transferred onto a fused silica substrate forming a stack of 5 graphene layers. The functionalization of as-grown graphene was accomplished by photoassisted hydrogenation reaction with formic acid, allowing increasing its nonlinear optical response and its laser-induced damage resilience.

\bibliographystyle{unsrt}  
\bibliography{references}  %%% Remove comment to use the external .bib file (using bibtex).

%\begin{thebibliography}{1}

%\bibitem{kour2014real}
%George Kour and Raid Saabne.
%\newblock Real-time segmentation of on-line handwritten arabic script.
%\newblock In {\em Frontiers in Handwriting Recognition (ICFHR), 2014 14th
%  International Conference on}, pages 417--422. IEEE, 2014.

%\bibitem{kour2014fast}
%George Kour and Raid Saabne.
%\newblock Fast classification of handwritten on-line arabic characters.
%\newblock In {\em Soft Computing and Pattern Recognition (SoCPaR), 2014 6th
%  International Conference of}, pages 312--318. IEEE, 2014.

%\bibitem{hadash2018estimate}
%Guy Hadash, Einat Kermany, Boaz Carmeli, Ofer Lavi, George Kour, and Alon
%  Jacovi.
%\newblock Estimate and replace: A novel approach to integrating deep neural
%  networks with existing applications.
%\newblock {\em arXiv preprint arXiv:1804.09028}, 2018.

%\end{thebibliography}

\end{document}